\documentclass{article}
\usepackage{ijcai22}

\usepackage[hidelinks]{hyperref}
\usepackage[small]{caption}
\usepackage[table]{xcolor}
\usepackage[utf8]{inputenc}
\usepackage{algorithmic}
\usepackage{algorithm}
\usepackage{amsfonts}
\usepackage{amsmath}
\usepackage{amssymb}
\usepackage{amsthm}
\usepackage{array}
\usepackage{arydshln}
\usepackage{booktabs}
\usepackage{colortbl}
\usepackage{float}
\usepackage{footmisc}
\usepackage{graphics}
\usepackage{graphicx}
\usepackage{latexsym}
\usepackage{makecell}
\usepackage{microtype}
\usepackage{multicol}
\usepackage{multirow}
\usepackage{soul}
\usepackage{tabularx}
\usepackage{times}
\usepackage{url}
\usepackage{xcolor}
\def\best{\bf\cellcolor[gray]{0.85}}
\def\secbest{\cellcolor[gray]{0.92}}

\newif\ifshowcomment
\showcommentfalse

\newcommand{\method}{\textsc{Sound2Synth}}
\ifshowcomment
    \newcommand{\todo}[1]{\textcolor{green}{[TODO: #1]}}
    \newcommand{\comm}[1]{\textcolor{gray}{[COMM: #1]}}
    
    \newcommand{\zui}[1]{\textcolor{red}{[Zui: #1]}}
    \newcommand{\yansen}[1]{\textcolor{blue}{[Yansen: #1]}}
\else
    \newcommand{\todo}[1]{}
    \newcommand{\comm}[1]{}
    
    \newcommand{\zui}[1]{}
    \newcommand{\yansen}[1]{}
\fi

\newtheorem{theorem}{Theorem}
\newtheorem{observation}{Observation}

\title{\method: Interpreting Sound via FM Synthesizer Parameters Estimation}

\author{
  Zui Chen$^{1\ast}$\and
  Yansen Jing$^{1\ast}$\and
  Shengcheng Yuan$^{2\ast}$\and
  Yifei Xu$^2$\and
  Jian Wu$^2$\And
  Hang Zhao$^1$\\
  \affiliations
  $^1$Institute for Interdisciplinary Information Sciences, Tsinghua University\\
  $^2$Beijing DeepMusic Technology Co., Ltd\\
  \emails
  \{chenzui19, jingys19\}@mails.tsinghua.edu.cn,
  \{yuansc,xuyf\}@lazycomposer.com,
  wujian7752@163.com,
  Zhaohang0124@gmail.com
}

\begin{document}

\maketitle
\renewcommand{\thefootnote}{\fnsymbol{footnote}}
\footnotetext[1]{These authors contributed equally to this paper.}
\footnotetext[2]{Code, plug-in, demo clips, demo video, and demo songs are all available at the link: \url{https://github.com/Sound2Synth/}.}
\renewcommand{\thefootnote}{\arabic{footnote}}

\begin{abstract}
Synthesizer is a type of electronic musical instrument that is now widely used in modern music production and sound design. Each parameter configuration of a synthesizer produces a unique timbre and can be viewed as a unique instrument. The problem of estimating a set of parameters configuration that best restore a sound timbre is an important yet complicated problem, i.e.: the synthesizer parameters estimation problem. We proposed a multi-modal deep-learning-based pipeline \method, together with a network structure Prime-Dilated Convolution (PDC) specially designed to solve this problem. Our method achieved not only SOTA but also the first real-world applicable results on the Dexed synthesizer, a popular FM synthesizer.
\end{abstract}

\section{Introduction}
\label{sec:introduction}

\subsection{Background and Motivation}
\label{sec:background-and-motivation}

The rapid development of computer hardware and digital audio processing technologies has greatly influenced the music industry in recent years. With the help of Digital Audio Workstation (DAW) and Virtual Studio Technology (VST), musicians are now able to finish the entire composition, orchestration, mixing, and mastering process on their computer or hardware for special use, which could provide real-time sound feedback to the musician, and are much cheaper than an actual band or professional orchestra.

As the virtual instruments become more delicate, they are also becoming larger in size -- professional virtual instrument libraries are reaching gigabyte level memory and terabyte level of disk storage for the audio samples -- and more expensive to record and use. Also, many sounds in musicians' imagination are not easily recordable or even producible in nature. Both problems inspired the idea of generating sound by only manipulating digital waveform signals, without the need of using a large size or a large number of samples.

A synthesizer is an electronic musical instrument that generates audio signals from given MIDI inputs, relying mainly on computation instead of audio samples. The main types of synthesizers include: additive/subtractive synthesizers, which construct waveforms by adding simple waves to silence or filtering out simple waves from white noises; FM synthesizers, which use cascaded oscillators whose parameters are controlled by predecessors to obtain complex waveforms at the output; wavetable-based synthesizers, which manipulate a short sample (i.e.: a wavetable) or combine multiple samples to create complex sounds; and analog synthesizers, which simulate physical circuits or devices to generate sounds.

Most synthesizers have a large number of parameters, which are necessary for their expressiveness. However, many of the parameters are non-intuitive for humans, which means that non-experts can not imagine a sound from a set of parameters configuration (i.e.: a {\it preset}), or conversely, determine a synthesizer preset from a given sound. This is the hard problem of Digital Sound Design, which is now evolving into a complex subject and major that requires years of learning and practicing to expertise in.

Our research addresses the specific problem of synthesizer parameters estimation in digital sound design: finding a close synthesizer preset given a sound. This is a problem that is not well-studied while having a huge potential benefit in terms of both industrial and artistic perspectives. For example:
\begin{itemize}
    \item {\bf Sound Design}: In digital sound design, musicians want to compose with natural/imaginal sounds as synthesizer presets (so that the sound can smoothly transfer to different pitches and durations). Synthesizer parameters estimation is able to remarkably speed up the process of making new presets, helping musicians to focus on the creative job only instead of the technical job.
    \item {\bf Sound Compression}: A synthesizer preset usually takes thousands of times or lesser space than the timbre sound wave it produces, thus synthesizer parameters estimation can be used in sound compression for efficient storage or transmission under circumstances that could tolerate minor trade-offs in quality for speed or memory efficiency (e.g.: online composing, preview rendering).
    \item {\bf Synthesizer Expressiveness Test}: A good synthesizer parameters estimator can also serve as a quantitative benchmark for comparing the expressiveness of synthesizers, by providing a set of random/naturally-distributed/domain-specific audio samples and evaluating the average distance between the original samples and the closest generated samples by each synthesizer.
\end{itemize}

In a nutshell, synthesizer parameters estimation is a problem of great importance to music creativity and has a great market value in the modern digital music industry.

\subsection{Problem Definition}
\label{sec:problem-definition}

Formally, a Synthesizer can be viewed as a function mapping $f:\Theta_f \times \mathcal M \to \mathcal A$, where $\mathcal A$ is the audio space, $\Theta_f$ is the configuration space of this particular synthesizer $f$, each $\theta \in \Theta_f$ is a preset of the synthesizer, $\mathcal M$ is the MIDI configuration space of a single note and each $\eta \in \mathcal M$ is the MIDI setting of a note, specified by note pitch, note velocity, and note duration, etc. Although this process is influenced by various other settings including sample rate, bit depth, etc, these settings are usually fixed and can be easily converted if necessary.

It is worth mentioning that almost all practical synthesizers are non-surjective functions, which means that there almost certainly exists audio $A \in \mathcal A$ that can not be generated by a specific synthesizer $f$.

The Synthesizer parameter estimation problem can be formulated as follows: given an audio $A \in \mathcal A$ and a fixed input note $\eta_0 \in \mathcal M$, the goal is to find a preset that can best restore the audio under a particular distance metric:%
\label{equation:definition}
\begin{equation}
\min_{\hat \theta \in \Theta_f} \text{dis}(f(\hat \theta,\eta_0),A)
\end{equation}

Our experiments mainly concentrate on the Dexed synthesizer~\cite{dexed}, since it is one of the most popular open-source FM synthesizers, and it has a sufficient amount of free presets easily accessible on the Internet. We will use mean squared Euclidean Distance between Mel-Frequency Cepstral Coefficients Distance (MFCCD) as our evaluation metric, since MFCCD is proven by experiment to be well-aligned with human perception~\cite{mfccd}, and is widely used in various audio-related tasks, e.g.: speech recognition.

\subsection{Related Work}
\label{sec:related-works}

The work in synthesizer parameters estimation can be traced back to the work of Horner et al.~\cite{horner1993machine}, which used a genetic algorithm to find the settings of parameters for frequency modulation matching synthesis. Mitchell et al. utilized the advances in multi-modal evolutionary optimization to perform dynamic sound matching of FM synthesizer~\cite{mitchell2005frequency}. 

Roth ~\cite{roth2011comparison} first introduced neural networks into this problem and systematically compared the performance of traditional methods and neural networks. Matthew Yee-King~\cite{vst-programming} further developed the usage of neural networks by using LSTM++ (bidirectional LSTM with highway connections) networks. They conducted experiments to compare their method APVST with various traditional approaches on the Dexed synthesizer, achieving comparable performance, while being much more efficient. However, results comparable to traditional approaches are still far from sufficient for real-world application.

After Yee-King's work, researches are also carried out on different synthesizers, e.g.: InverSynth~\cite{inver-synth} on JSyn~\cite{jsyn}, SerumRNN~\cite{serum-rnn} on Serum~\cite{serum}. Both works are using only simple network architectures: InverSynth used vanilla CNN networks, and SerumRNN used vanilla RNN networks. Also, Jsyn was a small synthesizer whose expressiveness is not sufficient for wide application, and SerumRNN focused only on the effects applied to the sounds, instead of the full preset configuration of Serum.

An interesting method FlowSynthesizer\cite{flow-synth} using auto-encoder and normalizing flow, was tested on Diva~\cite{diva} -- an analog synthesizer. This method does not rely on a running synthesizer instance but only on the data points (i.e.: audio paired with ground truth preset), and it achieved amazing results at the time. However, it was only able to achieve a relatively satisfactory result on a mere subset of Diva parameters instead of the entire Diva synthesizer. A similar method was used in follow by PresetGen VAE~\cite{le2021improving}, which is the previous SOTA on the Dexed synthesizer.

Another direction is to build differentiable synthesizers~\cite{ddsp}, which could then easily be integrated with neural network models. However, such a result could not be directly applied to existing commercial synthesizers, which are powerful but non-differentiable.

According to the works mentioned above, learning-based results comparable to traditional methods could be achieved, while there is no work so far producing human-tolerable results on a powerful synthesizer with full configuration space. Also, many of the works are limited to a single synthesizer or a synthesizer type.

Thus, we aim to develop a pipeline that could work on different types of synthesizers, achieving not only SOTA but also real-world applicable results on full configuration space.

\subsection{Novelty}
\label{sec:novelty}

Our main contributions in this paper are listed as follows:
\begin{itemize}
    \item We proposed {\bf \method} -- a deep-learning-based FM synthesizer parameter estimation pipeline, which does not require an online instance of a synthesizer during training. The proposed pipeline has achieved SOTA in quantitative comparison and is close to applicable in terms of human perception (Tab.~\ref{tab:exp}). Our pipeline could also in principle generalize to additive, subtractive and simple analog synthesizers.
    \item We proposed {\bf Prime-Dilated Convolution (PDC)} -- a new convolution network layer structure specially designed for better utilization of Constant-Q Transform (CQT) chromagram information of an audio sample.
    \item We demonstrated the benefit of using {\bf Multi-modal Features} in a combined network. The experiment showed that spectral (visual), waveform (sequential), and statistical (numerical) sound feature information altogether improved the generalizability of the model.
    \item We introduced various techniques for optimizing sound processing for neural networks and revealed many inspiring discoveries, which could as well be applied to other audio-related tasks.
\end{itemize}

\section{Methodology}
\label{sec:methodology}
\subsection{A Closer Look at the Dexed Synthesizer}

Frequency Modulation synthesis (or FM synthesis) is a form of sound synthesis whereby the frequency of an oscillator is changed by modulating its frequency with a modulator. The frequency of an oscillator is altered in accordance with the amplitude of a modulating signal. The modulator can be another oscillator, whose frequency can be modulated again by the third oscillator, and so on.

The Dexed synthesizer~\cite{dexed} is an open-source software synthesizer with 6 oscillators, aimed to model the FM synthesis algorithm in the Yamaha DX7 hardware, which is one of the best known and most successful synthesizers.

\begin{figure}[t]
    \centering
    \includegraphics[width=0.225\textwidth]{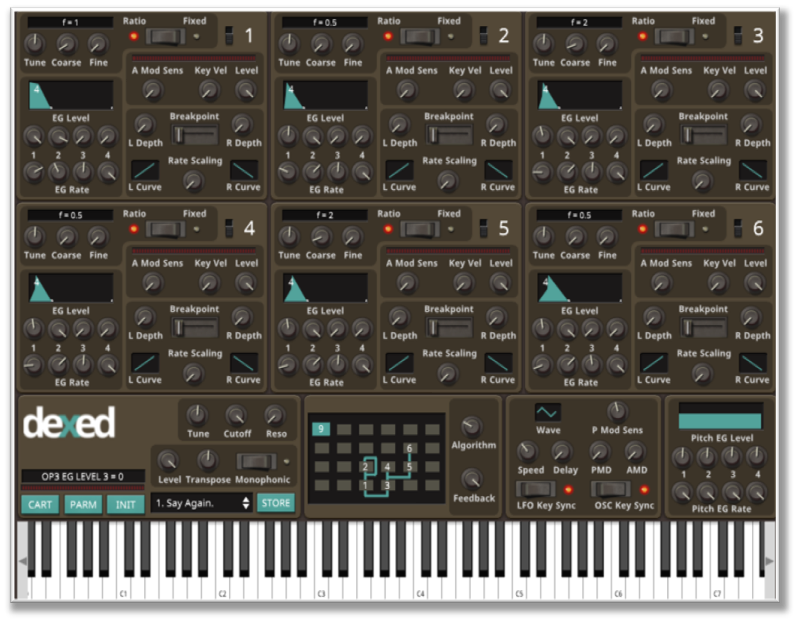}
    \includegraphics[width=0.232\textwidth]{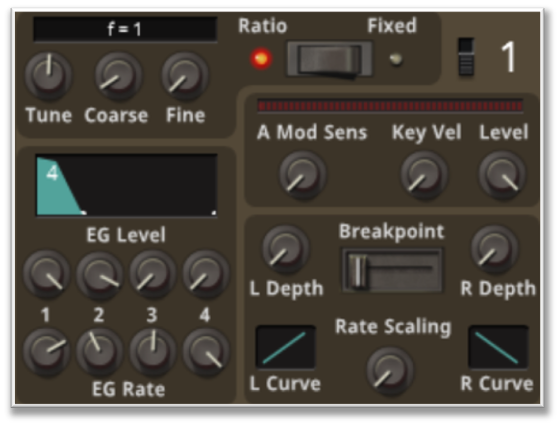}
    \caption{{\small {Screenshot of Dexed. Left: The user interface of the full Dexed synthesizer. Right: One oscillator of the Dexed synthesizer.}}
    \label{fig:dexed}}
\end{figure}

\subsection{Overall Pipeline}
\label{sec:overall-pipeline}

\begin{figure}[ht]
    \centering
    \includegraphics[width=0.47\textwidth]{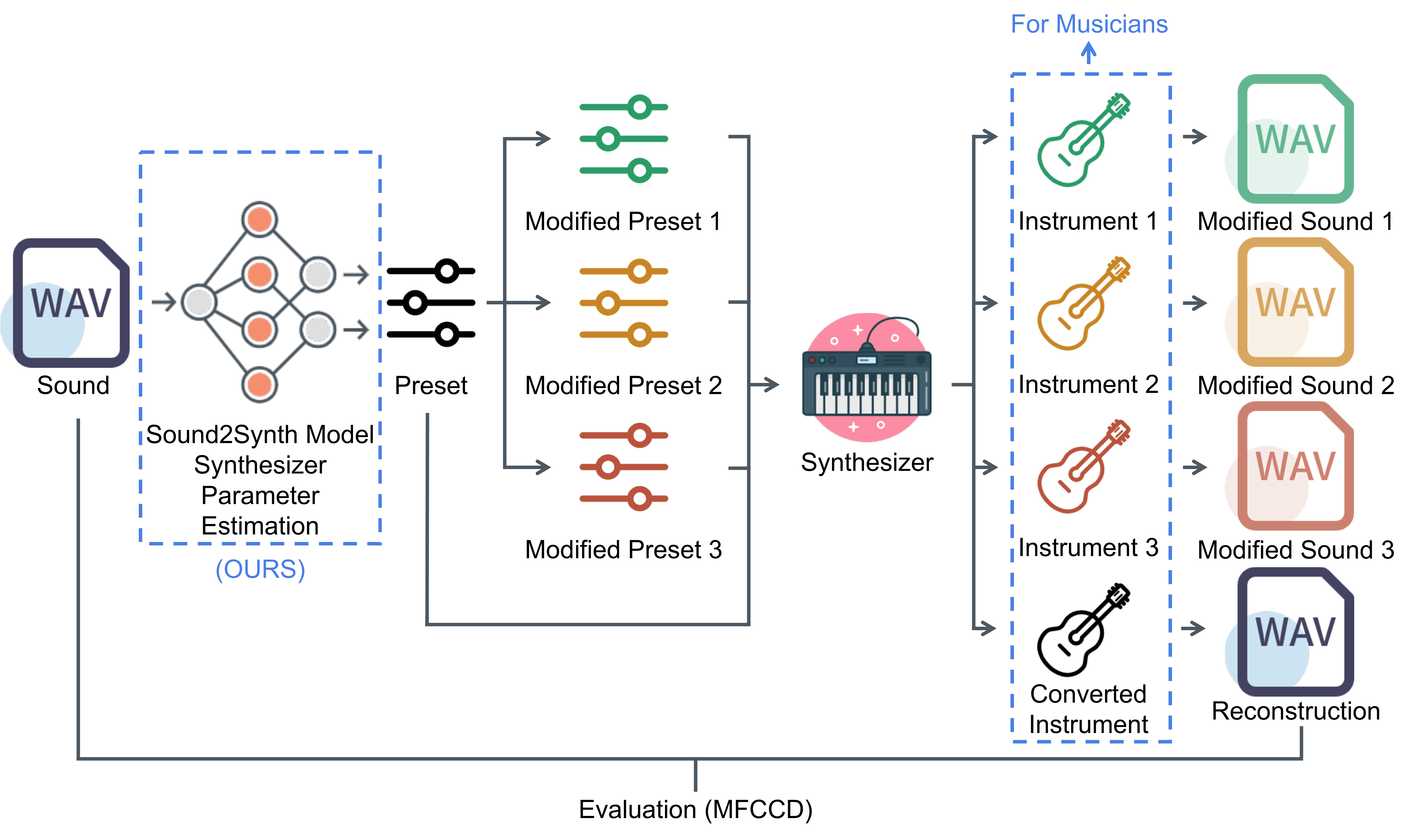}
    \caption{{\small {\method~overall pipeline illustration.}}
    \label{fig:workflow}}
\end{figure}
\begin{figure}[ht]
    \centering
    \includegraphics[width=0.49\textwidth]{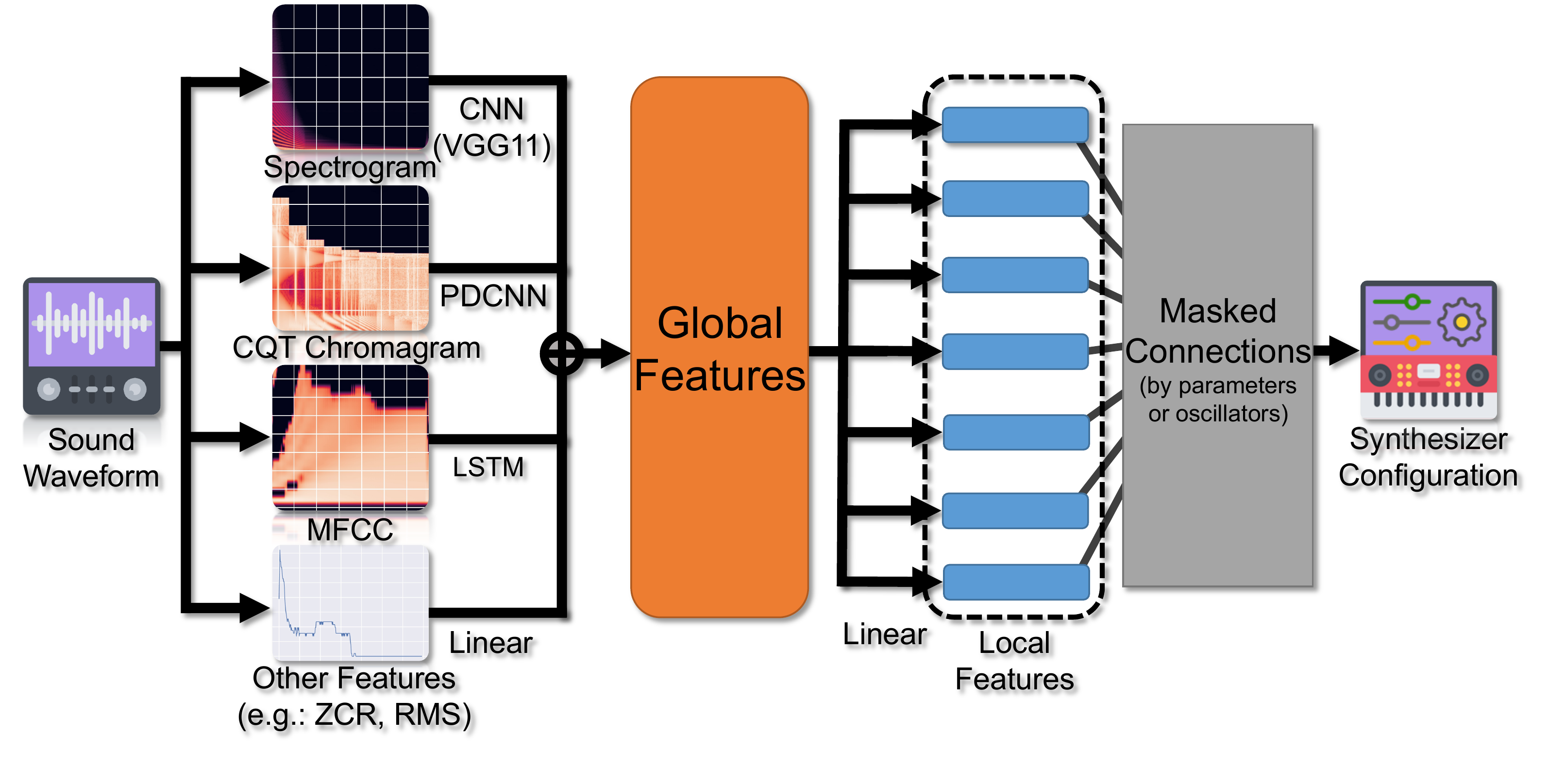}
    \caption{{\small {\method~model architecture illustration.}}
    \label{fig:method}}
\end{figure}

For each sound, it is first converted to different forms to address different aspects, including: Short-Time Fourier Transform (STFT) spectrogram, Mel spectrogram, CQT chromagram, MFCC, and other statistical information.

Each form of the input is then fed to the corresponding backbone that best fits for preprocessing: spectrograms are handled by CNNs\footnote{In fact, among vanilla CNN~\cite{lenet}, VGG~\cite{vgg}, ResNet~\cite{resnet} and DenseNet~\cite{densenet}, VGG prevailed to be the best backbone for audio-related tasks by preliminary experiments.}; chromagram is handled by Prime-Dilated Convolutional Neural Network (PDCNN), which is an original structure and will be elaborated in Sec.~\ref{sec:prime-dilated-convolution}; MFCC is handled by an LSTM network~\cite{lstm}; other statistical information is handled by simple MLPs. The processed features of each input form are concatenated to obtain the global features, containing all the information extracted from the input audio.

For parameters estimation, we first use a single linear layer with non-linearity to process the global features. Then we split the processed global features into groups of local features divided by oscillators or specific parameters. Then the local features are dealt with by multiple layers of linear and non-linearity connections masked according to parameters.

Finally, each parameter prediction is derived from the oscillator's local feature group using additional MLPs. Since continuous parameters in synthesizers usually have minimal precision, we can convert all continuous parameters into discrete parameters, therefore converting all real-value estimation problems to classification problems.

It is worth highlighting that, our pipeline supports not only online training but also training via offline datasets, which means it does not require access to a synthesizer instance during training. Synthesizer rendering is itself a computation-heavy operation, which would become the bottleneck if involved in the training procedure. Thus our workflow that allows separating the model from a synthesizer instance is more efficient and sometimes necessary in a real-world scenario.

Thus, we evaluate our model on configuration space $\Theta_f$ during training. That is if we denote the network as $N_{f,\varphi}:\mathcal A \to \Theta_f$, parameterized by $\varphi$, instead of optimizing the following true objective over an offline dataset $\mathcal D$:%
\label{equation:audio-mfcc-loss}
\begin{align}
  & \min_{\varphi} \mathcal L_{\text{MFCCD}}(\varphi) \nonumber\\
= & \min_{\varphi} \mathbb{E}_{(A^i,\theta^i) \sim \mathcal D}[ \text{MFCCD}(f(N_{f,\varphi}(A^i),\eta_0),A^i)]
\end{align}

During training, we only aim to optimize the Mean Squared Error (MSE) between predicted parameters and the groundtruth parameters:%
\label{equation:parameter-mse-loss}
\begin{equation}
\min_{\varphi} \mathcal L_{\text{MSE}}(\varphi) = \min_{\varphi} \mathbb{E}_{(A^i,\theta^i) \sim \mathcal D}[ \text{MSE}(N_{f,\varphi}(A^i),\theta^i)]
\end{equation}

This may cause dis-alignment between the training objective and the desired goal since:
\begin{itemize}
    \item Parameters have different importance, a wrongly predicted ``coarse'' parameter has a larger influence than a wrongly predicted ``fine'' parameter.
    \item Different configurations (in terms of numerical values of parameters) may produce similar sound timbres.
\end{itemize}

We proposed {\it gradient-inspired weighting} technique to handle this problem, which will be explained in Sec.~\ref{sec:gradient-inspired-weighting}.

\subsection{Prime Dilated Convolution}
\label{sec:prime-dilated-convolution}
\subsubsection{Intuition}
\begin{figure}[t]
    \centering
    \includegraphics[width=0.49\textwidth]{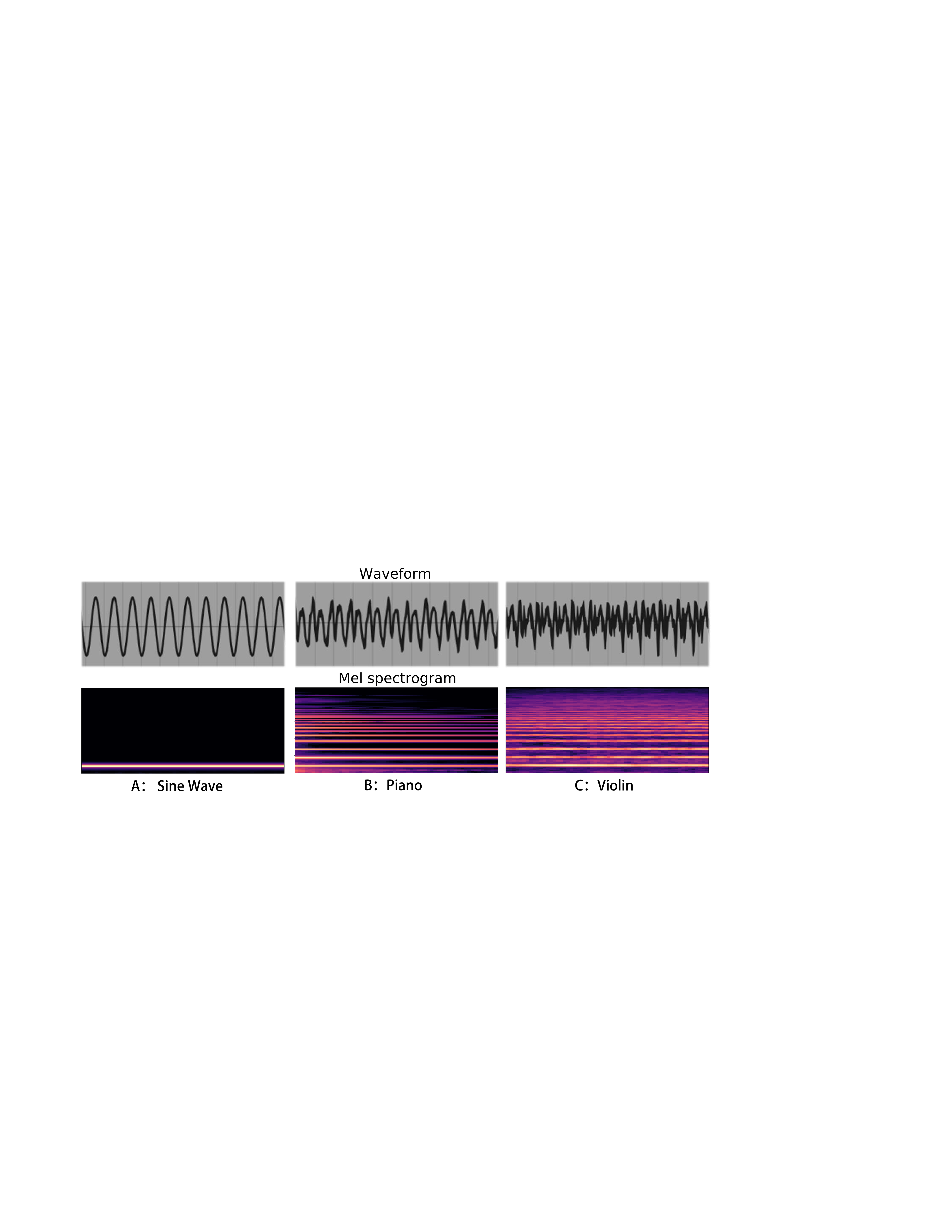}
    \caption{Illustration of different sound timbres having different harmonic features. From left to right, audios are obtained from sine wave, piano, and violin, separately. The audios are transformed into Mel-spectrogram. The phenomenon of multi-stripes shape due to the harmonics is very clear in these acoustic scenes.}
    \label{fig:Example}
\end{figure}

The general CNN treats image and spectrogram as the same data structure and ignores significant distinctions in their contents. One of the most essential distinctions comes from the harmonic characteristics of sound. 
Due to the physical properties of resonance frequency and mechanical waves, whenever a sound vibrates at fundamental frequency $F_0$, a series of frequencies at its integer multiples ($2F_0$, $3F_0$, etc.), called harmonics, are also likely to vibrate. This fact results in a common observation of multi-stripes shape vertically stacking in the spectrogram, as depicted in Fig.~\ref{fig:Example}. In this paper, we refer to this phenomenon as {\it harmonic features}.

Considering relationships among harmonics as different timbres, downstream networks (in this case, synthesizer parameters classifier) can utilize and leverage these harmonic features to improve performance. PDC constructs a sparse filter by expanding the concept of dilated convolution to accurately reach all the integer harmonics in a log-scale spectrogram. Unlike the regular dilated convolution which has a fixed dilated step, PDC's dilated location is not evenly distributed, since the distance between two harmonics in a spectrogram is not constant. PDC reaches all integer harmonics not at once, but through stacking itself. We apply the mathematical rule of prime factorization, decomposing an integer into the product of a few prime numbers. These primes are further decomposed into the product of some integer ratios between 1 and 2. PDC's dilated location is then built according to these ratios. In this way, PDC has a fixed receptive field, and it only requires a few primes to complete the filter construction, rather than traversing every integer.

\subsubsection{Prerequisite}
Let $\mathbf{X} \in \mathbb{R}^{C \times K \times T}$ denote the spectrogram, where $C$ is the number of channels, $K$ is the number of frequency bins, and $T$ is the number of temporal segments. 
Let $f(k)$ denote the frequency at the $k$-th bin, then $\mathbf{X}(c, k, t)$ refers to the energies of sound in $c$-th channel, at the $t$-th time frame, around frequency $f(k)$. The inverse function of $f$ is denoted as $f^{-1}(\cdot)$, which gives the index of the frequency bin according to the real frequency. 

The only prerequisite of PDC is for $\mathbf{X}$ to be a log-scale spectrogram, which means that $f(k)$ is an exponential function. In this paper, we use standard Constant-Q Transform (CQT) to generate spectrograms, where $f_{CQT}(k)=f_{\min} 2^{k/B}$, $f_{\min}$ is the lowest frequency that CQT covers (by default, $ f_{\min}\approx{32.70}{\text{Hz}}$), and $B$ denotes {\it bins per octave}, i.e., the number of bins between any frequency and its double frequency, which refers to the resolution of the spectrogram. 
The goal of prime-dilated convolution is to reach any integer harmonics in the spectrogram by setting the dilated location. Thus, there is a need to measure the distance between two harmonics. Let $d(n,m,F)=|f^{-1}(mF)-f^{-1}(nF)|$ denote the distance of bins between $n$-times and $m$-times harmonics based on their common fundamental frequency $F$. In CQT, the distance $d(n,m,F)$ does not change with $F$.%
\begin{align}
    &\forall{n,m\in \mathbb{N}},\quad \forall{F_1,F_2\in\mathbb{R}^{+}},\nonumber\\
    &d(n,m,F_1) = d(n,m,F_2) = \left| B\log_2\frac{m}{n} \right|
    \label{1}
\end{align}%
We will use $d(n,m)$ instead of $d(n,m,F)$ in the remaining part of the paper for simplicity.

Although $d(n,m)$ does not change with $F$, it is affected by $n$ and $m$, according to Eq.~\eqref{1}. Therefore, the dilated step cannot be constant. Because $n$ and $m$ may vary within a wide range, solving Eq.~\eqref{1} to get every possible distance between two harmonics would create a huge amount of dilated locations, i.e., creating a tall and dense filter with a lot of parameters. Instead, we introduce the prime-ratio function to represent all the distances using a small list of prime numbers and create a sparse filter.
\subsubsection{Prime-Ratio Function}

For any prime number $p$, the prime-ratio function $r(p)$ is defined as follows.%
\begin{align}
    r(p) &= 2^{-s}p,\quad s = \max\{s\in\mathbb{N}|2^s < p\}
    \label{2}
\end{align}%
Given the mathematical rule of prime factorization, i.e., for every $n\in\mathbb{N}$ and $n\ge 2$, there exists only one way to decompose $n$ into the product of prime powers. Considering that $r(2)=2$, therefore, $p=r(2)^s r(p)$, and the following theorem holds true for every integer.

\begin{theorem} For all $n \ge 2$, $n$ can be represented in exactly one way as a product of the prime-ratio powers, i.e.,%
\begin{align}
    n &= \prod_{i=1}^l r(p_i)^{\alpha_i}
    \label{3}
\end{align}%
where $p_1<p_2<\dots<p_l$ are prime numbers and $\alpha_i$'s are positive integers. 
\end{theorem}

Calculating $d(1,n)$ based on Eq.~\eqref{3} results in the following equation:%
\begin{align}
    d(1,n) &= \sum_{i=1}^l \alpha_i d(1,r(p_i))
    \label{4}
\end{align}%

Eq.~\eqref{4} states that the distance between any integer harmonic and its fundamental frequency can be represented as a finite linear summation of the prime-ratio's distance $d(1,r(p_i))$, where $d(1,r(p_i))$ has two characteristics:
\begin{enumerate}
    \item It always represents the distance between two integer harmonics, because according to Eq.~\eqref{1} and~\eqref{2},%
    \begin{align}
        d(1,r(p_i)) &= d(2^{s_i},p_i)
    \end{align}%
    where $2^{s_i}$ and $p_i$ are both integer numbers.
    
    \item It always lies between $0$ and $B$, since $r(p)\in (1,2]$ clearly is true for every prime, and $0=d(1,1)<d(1,r(p_i)) \le d(1,2) = B$.
\end{enumerate}

\subsubsection{Formulation}

\begin{figure}[t]
    \centering
    \includegraphics[scale=0.28]{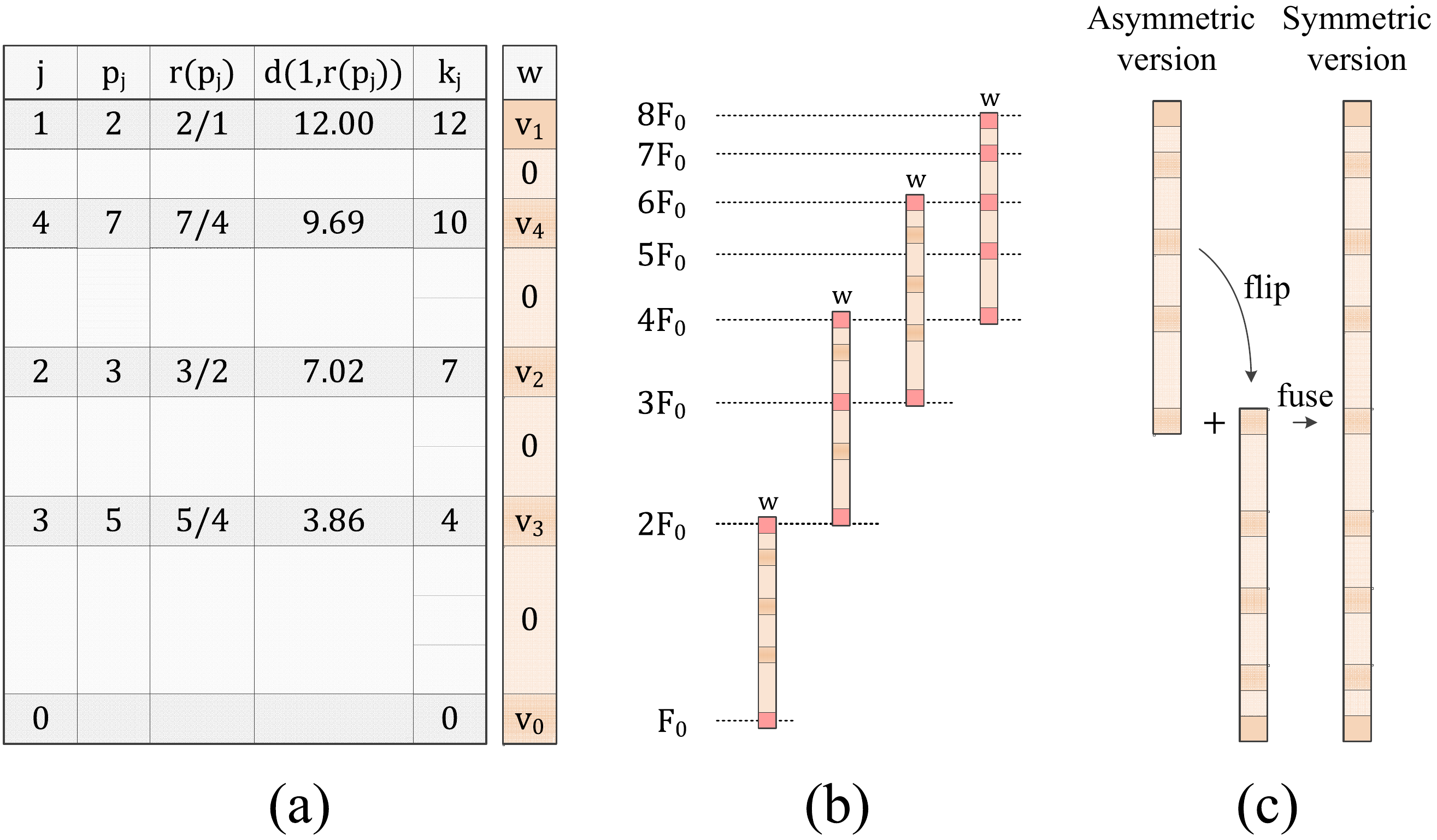}
    \caption{Main concept of the prime-dilated convolution, illustrated with the hyper-parameters $B=12$ and $l=4$. (a) shows the dilated structure of asymmetric version, where $j$ is index, $p_j$ is the $j$-th smallest prime, $r(p_j)$ is the prime-ratio function, $d(1,r(p_j))$ is the distance of bins to the bottom line, $\vec{v}$ is the trainable parameters, $\vec{w}$ is the filter after dilation, and $k_j$ is the dilated location. (b) illustrates the way the filter shifts and stacks to capture all the harmonics in a log-scale spectrogram. (c) introduces the symmetric version of PDC which is constructed from the first version's flip and fusion.
}
    \label{fig:PDC}
\end{figure}

The main concept of PDC is inspired by the above theorem and their inferences. Because: 1) any integer can be represented as the product of $r(p)$; 2) the product of $r(p)$ appears as the summation of $d(1,r(p))$; and 3) if the dilated location is set as $d(1,r(p))$, the summation refers to the shift and stack of convolution operation. In other words, if a series of distance $d(1,r(p_i))$ is taken as dilated locations as shown in Fig.~\ref{fig:PDC}, then any integer harmonics will be captured by the shift and stack of the filter. For example, $6=r(2)^2r(3)$, so the six-times harmonics can be reached by the stack of dilation $d(1,r(2))$, $d(1,r(2))$ and $d(1,r(3))$. In our experiments, we obtain such a stack by inserting a single PDC filter after every convolutional layer. All the distances $d(1,r(p))$ are no greater than $B$, leaving PDC with a fixed receptive field.

In practice, the dilated location in a filter has to be an integer, but $d(1,r(p_i))$ is often an irrational. Therefore, PDC constructs the dilation according to the integer approximation of $d(1,r(p_i))$ sequence, which is formally defined as follows:

Let $p_1<p_2<\dots<p_l$ denote the smallest $l$ prime numbers selected. Let $k_j$ denote the integer which is the closest to $d(1,r(p_j))$, we then have:%
\begin{align}
    k_0 &= 0,\nonumber\\
    k_j &= \arg \min_k | k - B\log_2 r(p_j) |,\quad j = 1,\dots,l
\end{align}%
We introduce two versions of the PDC filter in terms of how to construct the dilated locations.

\paragraph{Asymmetric version.} Let $K=\{k_j\}_{j=0}^{l}$ denote the set of dilated locations. Let vector $\vec{v} = (v_0,v_1,\dots,v_l)$ denote the trainable parameters in PDC. Let $\vec{w}=(w_0,w_1,\dots,w_B)$ denote the vector after dilation, where the receptive field is $(B+1) \times 1$. The values of $w_k$ are defined as follows. %
\begin{align}
    w_{k_j} &= v_j, \quad j=0,1,\dots,l \\
    w_k &= 0, \quad k \notin K \nonumber
    \label{7}
\end{align}%
Eq.~\eqref{7} creates an asymmetric structure $\vec{w}$ which covers only higher harmonics, as shown in Fig.~\ref{fig:PDC} (a) and (b).

\paragraph{Symmetric version.} Symmetric PDC is created by flipping and fusing the asymmetric version, as shown in Fig.~\ref{fig:PDC} (c). Let $K=\{k_j\}_{j=-l}^{l}$ denote the set of dilated locations, where the negative index of $k_{-j}$ is defined as $k_j$'s opposite:%
\begin{align}
    k_{-j} &= -k_j, \quad j=1,\dots,l
\end{align}%
Let vector $\vec{v} = (v_{-l},\dots,v_l)$ denote the trainable parameters in PDC, and $\vec{w}=(w_{-B},\dots,w_B)$ denote the vector after dilation, where the receptive field is $(2B+1) \times 1$. The values of $w_k$ are defined as follows:%
\begin{align}
    w_{k_j} &= v_j, \quad j=-l,\dots,l \\
    w_k &= 0, \quad k \notin K \nonumber
\end{align}%

The convolution operation ${pdc}(\cdot)$ can be parameterized with the dilated filter constructed by $\vec{w}$.

\subsection{Multi-modal Feature Engineering}
\label{sec:multi-modal-feature-engineering}

Besides spectrograms, chromagrams, and MFCC, we also utilized certain statistical features in the network, which are closely related to sound timbre. For example, the following information is widely used in audio processing tasks:
\begin{itemize}
    \item {\bf Amplitude Envelop}: The changes in the amplitude of a sound over time.
    \item {\bf RMS Energy}: The root mean square energy of audio.
    \item {\bf Zero Crossing Rate}: The rate at which a signal changes between positive value and negative value.
    \item {\bf Wiener Entropy}: Also known as {\it Spectral Flatness}, a metric to measure whether a sound is tonal or noisy.
\end{itemize}

Notice that the information is scalars per time step, given that a fixed input note has a fixed duration, we can directly use an MLP mapping from time steps to feature dimensions to process each statistical information.

\subsection{Techniques}
\label{sec:techniques}

\subsubsection{Label Smoothing}
\label{sec:label-smoothing}

As mentioned above, by discretizing continuous parameters, all parameter estimation could be treated as classification problems. In practice, all continuous parameters are in $[0,1]$ range and are divided into $K$ segments, as in a $K$-way classification task.

Unlike normal classification tasks, discretized numerical classes are not symmetric -- wrongly classifying a class as an adjacent class has a smaller influence than classifying it as an arbitrary other class. Thus, we can split part of the probability mass of the ground truth label into neighboring classes. Technically, the ground-truth label of length $K$ would first be 1d-convoluted with a Gaussian kernel of $\sigma = \sigma_0/K$, normalized to have a total probability mass sum to $1$, and then be used as the target for cross-entropy loss computation.

\subsubsection{Gradient-Inspired Weighting}
\label{sec:gradient-inspired-weighting}

As mentioned in Eq.~\eqref{equation:parameter-mse-loss}, only considering MSE loss on parameter space could result in overfitting the configuration space, while performing badly in the audio space.

\begin{observation}
Most parameters in most presets are local continuous: a small change in the preset would also indicate a small change in the rendered audio and MFCC.
\end{observation}%
This implies that we can approximate local audio space loss using a linear loss term.

\begin{observation}
Based on preliminary experiments, our model would be able to generate predictions relatively close to the ground truth.
\end{observation}%
This implies that we can use the gradient field around ground truth $\theta^\ast$ to substitute the one around prediction $\hat \theta$.

Combining the observations, we can approximately state:
\begin{align}
  & \left.\frac{\partial \mathcal L_{\text{MFCCD}}(\theta)}{\partial \theta_i}\right\vert_{\theta=\hat\theta} \nonumber\\
= & \left.\frac{\partial \mathcal L_{\text{MFCCD}}(\theta)}{\partial \mathcal L_{\text{MSE}}(\theta_i)}\right\vert_{\theta=\hat\theta}\cdot\left.\frac{\partial \mathcal L_{\text{MSE}}(\theta_i)}{\partial \theta_i}\right\vert_{\theta=\hat\theta} \nonumber\\
\approx & \left.\frac{\Delta \mathcal L_{\text{MFCCD}}(\theta)}{\Delta \mathcal L_{\text{MSE}}(\theta_i)}\right\vert_{\theta=\theta^\ast}\cdot\left.\frac{\partial \mathcal L_{\text{MSE}}(\theta_i)}{\partial \theta_i}\right\vert_{\theta=\hat\theta}
\end{align}

By preprocessing $\left.\frac{\Delta \mathcal L_{\text{MFCCD}}(\theta)}{\Delta \mathcal L_{\text{MSE}}(\theta_i)}\right\vert_{\theta=\theta^\ast}$ for each parameter $\theta_i$ of each preset $\theta^\ast$ in dataset $\mathcal D$, we can estimate an importance weight of prediction $\hat \theta_i$, to be used in training.

This technique could be applied only if the number of training samples in dataset $\mathcal D$ is small, since rendering audio using a synthesizer and computing audio space loss for every parameter of every data point is very time-consuming. Optimizing this method will be left as future work.

\section{Dataset}
\label{sec:dataset}

There are $155$ parameters for the Dexed synthesizer in total: $87$ continuous parameters, $66$ discrete parameters, and $2$ fixed parameters ({\it Algorithm} and {\it Output}).
\begin{figure}[t]
    \centering
    \includegraphics[width=0.45\textwidth]{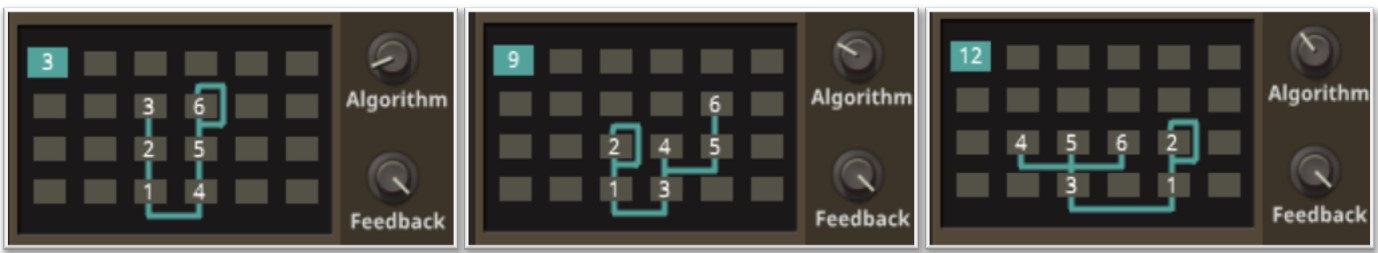}
    \caption{{\small {Different {\it Algorithm}s. There are 32 {\it Algorithm}s in total.}}
    \label{fig:algorithm}}
\end{figure}

{\it Algorithm} is a special parameter in Dexed, which determines the modulation relationship between the six oscillators (Fig.~\ref{fig:algorithm}). The physical meaning of all parameters depends on the choice of {\it Algorithm}. WLOG, we restrict our experiment on a fixed {\it Algorithm} setting. In practical application, a model should be trained for different algorithms respectively. And then we select the output of the model with the smallest audio space loss as the final output of the system.

There is another special parameter called {\it Output}, which controls the volume of the generated sound. In our datasets, we fixed the value of {\it Output} to the maximum value, otherwise, the generated audio loudness of many presets is so low that it affects the perception of the human ear. In the model training and inferencing process, we also fixed the {\it Output} parameter, which is consistent with the data set. Fixing this parameter is also a reasonable decision in practice since users can easily adjust the overall volume of the synthesizer output afterward, and it is often necessary to do so.

We combined three different methods to generate datasets:
\begin{itemize}
    \item {\bf Preset Based}: We collected 100k+ presets on the Internet, which are widely used in real-life music composition. We input each preset $\theta^i$ into the Dexed synthesizer to render the corresponding audio files $A^i$. We split the preset-audio pairs into $32$ sub-datasets according to different values of {\it Algorithm}.
    \item {\bf Preset Augmentation}: We applied simple data augmentation to enlarge the dataset. Given a preset, we can fix the value of most of its parameters and then uniformly sample the values of the remaining parameters.\footnote{Dexed presets are grouped into different {\it themes}, which are split and augmented separately so that there is no data leakage during augmentation.} Note that randomly generated presets may not be audible, we set a minimum threshold of audible volume to sieve those presets out.
    \item {\bf Random Walk Based}: To improve and test generalizability, besides collecting presets online, we also randomly sampled presets from configuration space to construct a random dataset. Similarly, we only preserved presets that can generate audible sounds.
\end{itemize}

\section{Results}

\newcolumntype{C}[1]{>{\centering\arraybackslash}m{#1}}
\begin{table}[H]
\small
\centering
\renewcommand{\arraystretch}{0.9}
\resizebox{0.98\linewidth}{!}{
\begin{tabularx}{0.86\linewidth}{@{}m{0.6\linewidth}C{0.18\linewidth}}
\multicolumn{2}{c}{\textbf{Quantitative Results on Dexed}} \\ \toprule
{\bf Method} & {\bf MFCCD} \\ \midrule
*Hill Climbing & 21.96 \\
*Genetic Algorithm & 31.32 \\
APVST MLP & 31.38 \\
APVST LSTM & 32.76 \\
APVST LSTM++ & 22.59 \\
PresetGen VAE & 14.70 \\
\midrule
*Similarity Threshold & \multirow{2}{*}{10 $\sim$ 15} \\
for Human Perception & \\
\midrule
\method~{\tiny\textcolor{blue}{(OURS)}} & {\secbest 6.32} \\
\method~multi-modal {\tiny\textcolor{blue}{(OURS)}} & {\best 5.36} \\
\bottomrule
\end{tabularx}
}
\caption{{\small Experiment results. MFCCD is the lower the better. All MFCCDs are measured under T6 setting: 6 oscillators on Dexed.}}
\label{tab:exp}
\end{table}
\begin{figure}[ht]
    \centering
    \includegraphics[width=0.48\textwidth]{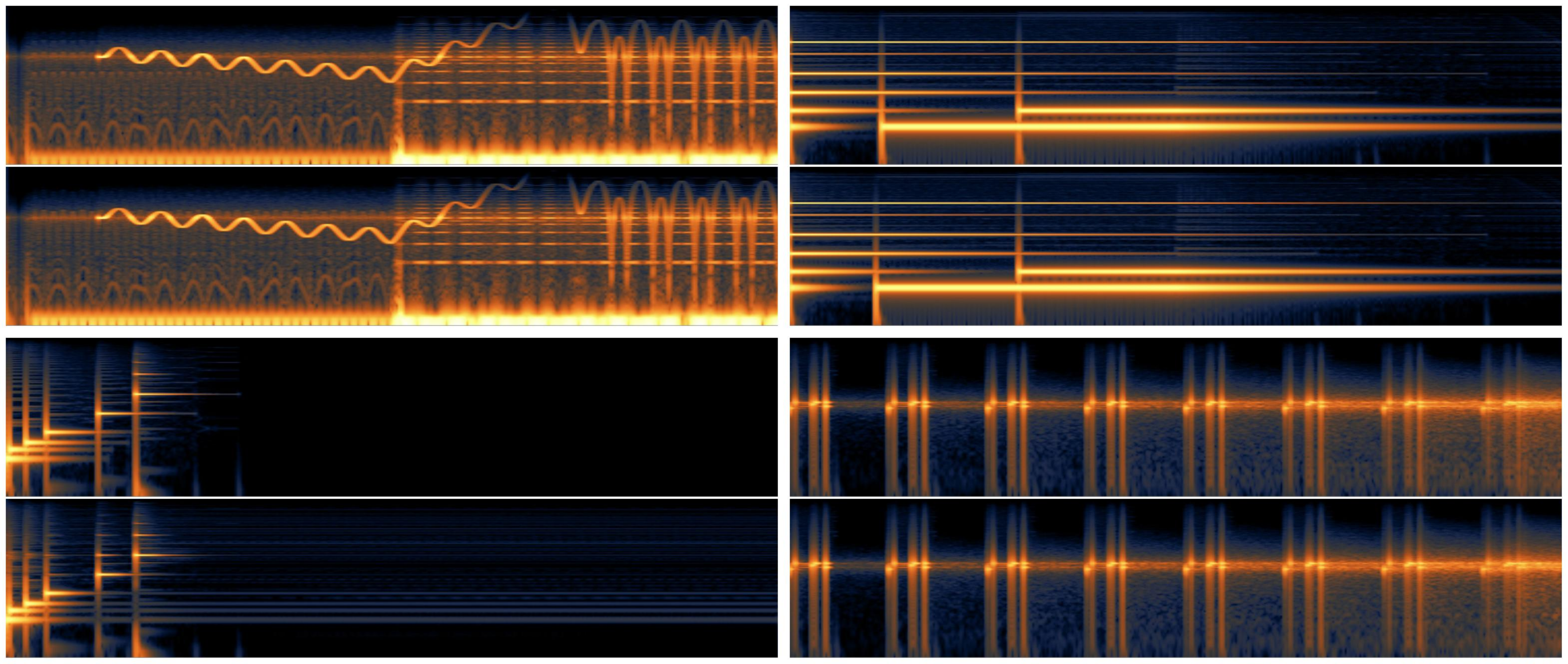}
    \caption{{\small Four sampled spectrogram cases. Ground truth audios are on the top and audios generated using predicted parameters are on the bottom inside each group.}}
    \label{fig:case}
\end{figure}
\renewcommand{\thefootnote}{\fnsymbol{footnote}}
\footnotetext[1]{These figures obtained from APVST~\cite{vst-programming}. Our subjective listening test also agrees with this similarity threshold.}
\renewcommand{\thefootnote}{\arabic{footnote}}

The detailed experiment settings are elaborated in Appendix.~\ref{sec:exp-set}.

From a quantitative perspective (Tab.~\ref{tab:exp}), our model largely outperforms previous SOTA: PresetGen VAE~\cite{le2021improving}.\footnote{MFCCD metric is influenced by the number of filter banks, however, we computed MFCCD under both 13-band (\cite{vst-programming}) and 40-band (\cite{le2021improving}) settings and observed no remarkable difference in the evaluation results. Thus we reported the 13-band MFCCD in Tab.~\ref{tab:exp}.} From a visual perspective (Fig.~\ref{fig:case}), the spectrograms of our predictions are very similar to that of the ground truths. From an auditory perspective, audios generated using predicted preset and the ground truth audio are very alike.

\section{Conclusion}
\label{sec:conclusion}
We proposed a novel multi-modal pipeline, along with a prime-dilated convolution structure and many other useful techniques in audio processing, to tackle the synthesizer parameters estimation problem. The result of our pipeline, \method, is not only significantly better than previous SOTA on the Dexed synthesizer but also able to reach human auditory perception precision. We have released code, plug-in, audio demos, and example use cases in which our plug-in is boosting musicians' creativity and simplifying the process of creation substantially. This could have an impact on the development of AI for art in the field of music and sound design, and it could be beneficial for other audio processing tasks in the future.

\appendix
\section{Experiment Settings}
\label{sec:exp-set}

WLOG, we fixed the input note $\eta_0$ to be at the middle C pitch (C4). The note is always pressed with maximum velocity, sustained for $4$ beats, and recorded $8$ beats in total, under tempo $120$ bpm. All audios are converted to $48\text{kHz}$ sample rate and $32$ bit depth. All $6$ oscillators of Dexed are used, including $155$ parameters in total: $87$ continuous parameters, $66$ classification parameters, and $2$ fixed parameters. All continuous parameters are discretized into $64$ classes.

Our experiments are carried out on a pre-generated dataset containing $30106$ training/validation data points and $1679$ test data on Dexed. Among the training/validation data points, $6191$ are directly sampled from existing presets, $22237$ are augmented from those presets, and $1678$ are generated purely at random. In practice, $80\%$ of the data points are used for training and $20\%$ are held out for validation. Notice that the test dataset is generated from independent held-out themes of presets and random-walk is not used, preventing data leakage.

On model architecture, the extracted global features have the same dimension of $2048$ for all model structures. In the case of the multi-modal structure, each backbone is assigned a small portion of features. Specifically, convolutional backbones, which are used to extract features from spectrogram and CQT chromagram, each have an output dimension of $512$, while other backbones, which are used to extract features from waveform, MFCC, or statistical information, each have an output dimension of $128$. The masked classifier has $64$ hidden neurons for each group (a parameter or an oscillator).

We trained our models using the AdamW~\cite{adamw} optimizer with a universal weight decay $10^{-4}$ and a linear warm-up cosine annealing scheduler with $4$ fixed warm-up epochs and a peak learning rate $2 \times 10^{-4}$ over at most $30$ epochs. We used a virtual batch size of $64$ data points per batch by using gradient accumulation. We adopted training tricks including gradient clipping, snapshot, early stopping, stochastic weight averaging, etc. It is worth noticing that small Gaussian noise is added to training data points to improve the robustness of the model.

We trained each of our models on a Linux server using a single NVIDIA GeForce GTX 1080Ti GPU. The maximum GPU RAM usage is no more than $9$GB for a properly chosen physical batch size.

\section{\method~Plug-In}
\label{sec:plug-in}

\begin{figure}[ht]
    \centering
    \includegraphics[width=0.44\textwidth]{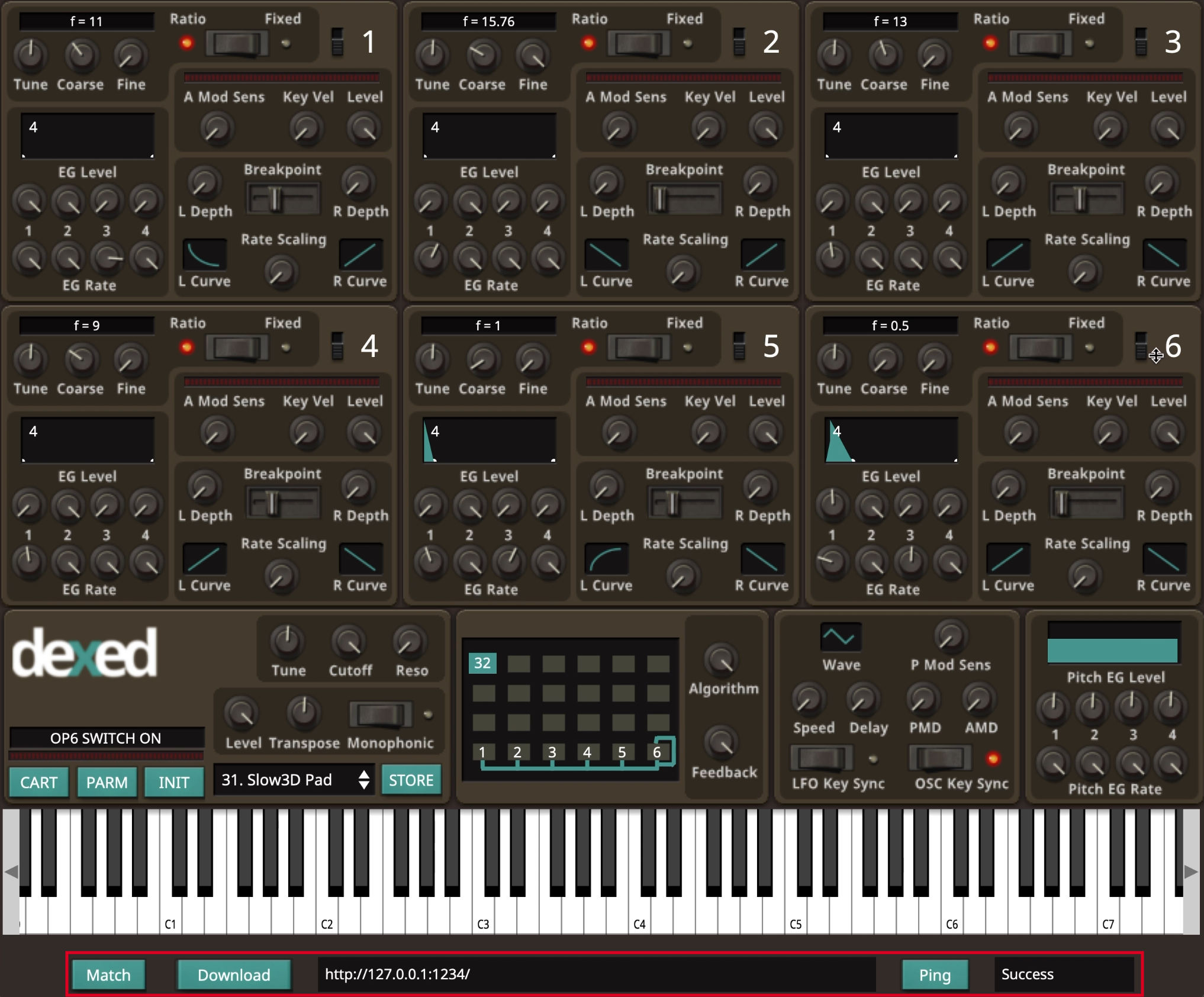}
    \caption{{\small {Screenshot of \method~plug-in built on Dexed. The part highlighted by the rectangle is the \method~interface.}}
    \label{fig:plug-in}}
\end{figure}

Using our \method~model, we developed and released a plugin based on the Dexed synthesizer. The plug-in first ``Ping'' the server running the neural network to establish a connection. Then by ``Match''ing an input audio file, our \method~model will automatically calculate the corresponding parameters and assign them back to the synthesizer. The plug-in also supports ``Download'' to serialize and save preset in human-readable JSON format.

\section*{Ethical Statement}

There are no ethical issues.

\section*{Acknowledgments}

We would like to thank Xiaoguang Liu from Beijing DeepMusic Technology Co., Ltd for providing computing resources and managing the project. We would also like to thank Yang Yuan and Jiaye Teng from IIIS, Tsinghua University for supporting this research.

\bibliographystyle{named}
\bibliography{ijcai22}

\end{document}